\documentclass[10pt,twocolumn,aps,prl,superscriptaddress]{revtex4-2}
\usepackage[utf8]{inputenc}
\usepackage{booktabs}
\usepackage{amsmath}
\usepackage{amsthm}
\usepackage{amssymb}
\usepackage{graphicx}
\usepackage{hyperref}

\makeatletter

\providecommand{\tabularnewline}{\\}

    \usepackage{physics}
    \usepackage{amsthm}
\usepackage{siunitx}
    \usepackage{tikz}

    \newcommand*{\balancecolsandclearpage}{%
        \close@column@grid
        \cleardoublepage
        \twocolumngrid
    }
    \newcommand{\cmmnt}[1]{}

\makeatother

\begin{document}
\title{Time-optimal Qubit Reset via Environmental Spectral Structure}
\author{Hong-Bo Huang}
\affiliation{Graduate School of China Academy of Engineering Physics, No.10 Xibeiwang
East Road, Haidian District, Beijing, 100193, China}
\author{Hui Dong}
\email{hdong@gscaep.ac.cn}

\affiliation{Graduate School of China Academy of Engineering Physics, No.10 Xibeiwang
East Road, Haidian District, Beijing, 100193, China}
\date{\today}
\begin{abstract}
Fast qubit reset is essential for qubit reuse in the noisy intermediate-scale quantum computing era, yet it conflicts with the weak decoherence required for
high-fidelity computation. We solve the time-optimal reset problem
for a frequency-tunable qubit coupled to a structural environment
under realistic spectral and control constraints. The optimal strategy
consists of a switch--restore--switch
sequence, where the qubit is moved from a low-decoherence
computational configuration to a high-decoherence restoring configuration
and then returned for reuse. For superconducting qubits in four representative
environments, this strategy reduces the reset time from typically
$\gtrsim\SI{100}{\nano\second}$ to $\SI{20}{\nano\second}$,
about $40\%$ of a typical two-qubit gate time, while achieving a
reset precision of $10^{-5}$. Our results identify environmental
spectral structure as a practical resource for rapid, high-fidelity
qubit reset and provide a design principle for qubit reuse on qubit-limited
processors.
\end{abstract}
\maketitle
Many quantum algorithms require more logical qubits than are available
in the noisy intermediate-scale quantum (NISQ) computing era \citep{barenco1995elementary,divincenzo2000physical,preskill2018quantum,decross2023qubit,abughanem2025ibm}.
Qubit reuse via mid-circuit reset is therefore essential for extending
circuit depth and executing larger algorithms with limited hardware
resources \citep{divincenzo2000physical,chow2009randomized,reed2010fast,geerlings2013demonstrating,zhou2021rapid,hua2022exploiting,mcewen2021removing,decross2023qubit,hu2024engineering,abughanem2025ibm}.
In this context, reset is not a peripheral operation but an integral
part of the computational cycle itself, whose duration directly sets
the qubit recycling rate and can become a bottleneck for computational
throughput \citep{divincenzo2000physical,chow2009randomized,reed2010fast,tsai2010toward,oliver2013materials,geerlings2013demonstrating,kjaergaard2020superconducting,zhou2021rapid,mcewen2021removing,decross2023qubit,hu2024engineering,jiang2025advancements,naeij2025open}.
Fast, high-fidelity reset is thus a prerequisite for scalable quantum
computation on limited hardware.

The central difficulty in achieving such rapid reset arises from a hardware-level tension between
computation and reset \citep{oliver2013materials,kjaergaard2020superconducting,zhou2021rapid,maurya2024demand,ding2025multipurpose,naeij2025open},
where high-fidelity gate operations require weak decoherence \citep{chow2009randomized,tsai2010toward,oliver2013materials,gu2017microwave,kjaergaard2020superconducting,jiang2025advancements,naeij2025open},
whereas rapid reset requires strong decoherence \citep{maurya2024demand}.
Passive reset via the natural energy relaxation is typically slow,
requiring on the order of $5\sim10T_{1}$ \citep{zhou2021rapid,yoshioka2023active,hu2024engineering}, where $T_{1}$ is the coherence time and always far longer than typical two-qubit
gate times for the superconducting qubit (SC-qubit) \citep{wallraff2005approaching,tsai2010toward,peterer2015coherence,kjaergaard2020superconducting,sah2024decay,jiang2025advancements,abughanem2025ibm,chow2009randomized}.
This mismatch motivates active reset schemes via the structural environment \citep{breuer2002theory,oliver2013materials,wang2021nonadiabatic,naeij2025open}, where the qubit is switched
from a low-decoherence computational configuration to a high-decoherence
reset configuration \citep{reed2010fast,zhou2021rapid,mcewen2021removing,hu2024engineering,maurya2024demand,ding2025multipurpose}.
Experiments have demonstrated substantial acceleration along these lines,
achieving reset times of order $\gtrsim\SI{100}{\nano\second}$ \citep{reed2010fast,magnard2018fast,mcewen2021removing,zhou2021rapid,ding2025multipurpose}. However,
the time-optimal reset strategy under realistic spectral and control
constraints remains unclear.

\begin{figure}
\includegraphics{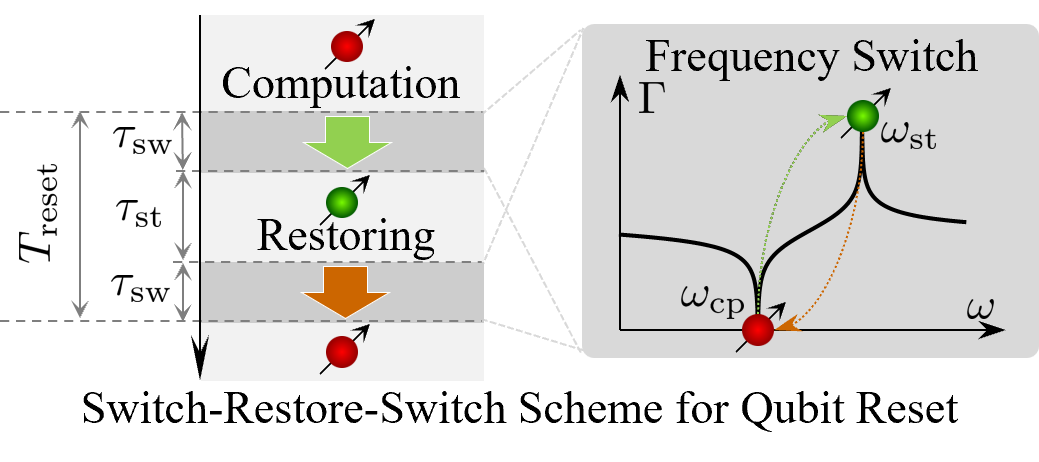} \caption{Illustration of the switch--restore--switch scheme for qubit reset using
the environmental spectral structure. The qubit exhibits a frequency-dependent
decoherence rate, $\Gamma\left(\omega\right)$. Initially, to perform
a computational task, the qubit operates in the computational configuration
(red) at the computation frequency $\omega_{\text{cp}}$ with low decoherence,
$\Gamma\left(\omega_{\text{cp}}\right)$. After the computational
task, it is switched over the switching duration $\tau_{\text{sw}}$
to the restoring configuration (green) at the restoring frequency
$\omega_{\text{st}}$ with rapid decoherence, $\Gamma\left(\omega_{\text{st}}\right)$.
In this configuration, the qubit undergoes strong decoherence and
relaxes over the restoring duration, $\tau_{\text{st}}$. Finally,
it is switched back to the computational configuration for reuse. }\label{FigIllustration}
\end{figure}

In this Letter, we determine the time-optimal reset scheme for a frequency-tunable
qubit coupled to a structural environment. As illustrated in FIG.~\ref{FigIllustration},
the optimal strategy consists of a switch--restore--switch
sequence. During computation (red qubit in FIG.~\ref{FigIllustration}),
the qubit typically operates at the computation frequency $\omega_{\text{cp}}$
with a long coherence time, $T_{\text{1}}=1/\Gamma\left(\omega_{\text{cp}}\right)$,
where $\Gamma\left(\omega\right)$ denotes the frequency-dependent
decoherence rate set by the spectral structure of environment-qubit
coupling (the right side of FIG.~\ref{FigIllustration}). In frequency-tunable SC-qubit, $T_{\text{1}}$ typically reaches $\SI{100}{\micro\second}$ \citep{wallraff2005approaching,tsai2010toward,peterer2015coherence,kjaergaard2020superconducting,sah2024decay,jiang2025advancements,abughanem2025ibm,chow2009randomized}. Upon completion
of a computational task, the qubit is switched to the restoring frequency
$\omega_{\text{st}}$ (green qubit in FIG.~\ref{FigIllustration})
over a switch duration, $\tau_{\text{sw}}$. The restoring duration,
$\tau_{\text{st}}$, is then related to the short coherence time,
$1/\Gamma\left(\omega_{\text{st}}\right)$. After restoring, the qubit
is switched back to the computational configuration for subsequent
operation. In this scheme, the total reset time is $T_{\text{reset}}=2\tau_{\text{sw}}+\tau_{\text{st}}$.
The switching duration, $\tau_{\text{sw}}$, is generally fixed by
the qubit design and is typically of order $\tau_{\text{sw}}\sim\SI{10}{\nano\second}$
for SC-qubits \citep{sandberg2008tuning,dicarlo2009demonstration,reed2010fast,pechal2016superconducting,mcewen2021removing,zhou2021rapid}.
Therefore, the restoring duration $\tau_{\text{st}}$ is the primary
target for reducing the reset time.

The restoring duration $\tau_{\text{st}}$ is determined by the qubit
dynamics during reset, which are governed by the Lindblad master equation, $\dot{\rho}_{S}=-\mathrm{i}\left[H_{S},\rho_{S}\right]+\mathcal{L}\left[\rho_{S}\right]$
\citep{breuer2002theory,wang2021nonadiabatic,naeij2025open}, where
$\rho_{S}$ is the reduced density matrix of the qubit, and $H_{S}=\omega\left(t\right)\sigma_{z}/2$
is the Hamiltonian for a frequency-tunable qubit with $\omega\left(t\right)$
denoting the instantaneous qubit frequency. The explicit form of the
Lindbladian superoperator $\mathcal{L}$ is provided in the End Matter
\citep{EndMatter}. The corresponding dynamics of the excited-state
population, $p_{e}$, is derived as \citep{EndMatter}
\begin{align}
\dot{p}_{e}= & -\Gamma\left(\omega\right)\left(p_{e}-p_{e}^{\text{eq}}\left(\omega\right)\right)\text{,}\label{EqEntropySt}
\end{align}
where $p_{e}^{\text{eq}}\left(\omega\right)=1/2\left(1-\tanh\left(\beta\omega/2\right)\right)$
is the thermal-equilibrium excited-state population at frequency $\omega$,
where $\beta=1/k_{B}T$ is the inverse environment temperature. The
frequency-dependent rate, $\Gamma\left(\omega\right)$, reflects the
spectral structure of the qubit-environment coupling.

We consider a restoring process of the qubit with maximum entropy.
Such a process starts from the qubit's initial state $p_{e}\left(0\right)=0.5$
and restores the qubit to its ground state within a reset precision,
i.e., $p_{e}\left(\tau_{\text{st}}\right)=\epsilon$. The qubit frequency
is typically subject to a path constraint $\omega\left(t\right)\in\left[\omega^{\text{min}},\omega^{\text{max}}\right]$
for all $t$, where $\omega^{\text{min}},\omega^{\text{max}}$ are
set by the control hardware. For SC-qubits, frequency tuning is commonly
achieved by adjusting the magnetic flux through a dc-SQUID loop\citep{gu2017microwave},
which imposes the path constraint generally written as $\omega^{\text{max/min}}=\omega_{\text{cp}}\pm\Delta\omega$.
The typical parameters are $\omega_{\text{cp}}\sim2\pi\times\SI{5}{\giga\hertz}$
\citep{houck2008controlling,bronn2015reducing,yan2016flux,gu2017microwave,bronn2017fast,kjaergaard2020superconducting,sah2024decay,jiang2024situ,jiang2025advancements,abughanem2025ibm,ding2025multipurpose}
and $\Delta\omega\sim2\pi\times\SI{3}{\giga\hertz}$ \citep{houck2008controlling,bronn2015reducing,yan2016flux,bronn2017fast,sah2024decay,ding2025multipurpose}.
The upper frequency bound sets the achievable reset precision, $\epsilon^{\text{min}}=p_{e}^{\text{eq}}\left(\omega^{\text{max}}\right)$.

Given the path constraint and the equation of motion in Eq. (\ref{EqEntropySt}),
minimizing the restoring duration, $\tau_{\text{st}}=\int_{0}^{\tau_{\text{st}}}\mathrm{d}t$,
constitutes an optimal control problem in which the tunable frequency,
$\omega\left(t\right)$, serves as the control field. Applying Pontryagin's
minimum principle \citep{pontryagin1987mathematical,EndMatter}, we construct
the control Hamiltonian \citep{EndMatter} as $\mathcal{H}\left(p_{e},\lambda;\omega\right)=1-\lambda\Gamma\left(\omega\right)\left(p_{e}-p_{e}^{\text{eq}}\left(\omega\right)\right)$,
where $\lambda$ is the Lagrange multiplier to incorporate the dynamical
equation, Eq. (\ref{EqEntropySt}). Since $\mathcal{H}$ does not depend
on the derivative of the control field, the optimization becomes time-local
\citep{albarelli2018locally,predko2020time}. Such a time-local optimal
control $\omega^{*}=\arg\min_{\omega\in\left[\omega^{\text{min}},\omega^{\text{max}}\right]}\mathcal{H}$
can be chosen independently at each instant, reducing the original
functional optimization to a pointwise static optimization. With $\lambda\left(t\right)>0$
for all $t$ \citep{EndMatter}, the optimal restoring trajectory
satisfies a local algebraic relationship
\begin{align}
\omega^{*}\left(p_{e}\right)= & \arg\max_{\omega\in\left[\omega^{\text{min}},\omega^{\text{max}}\right]}\Gamma\left(\omega\right)\left(p_{e}-p_{e}^{\text{eq}}\left(\omega\right)\right)\text{.}\label{EqAlgRela}
\end{align}
This relation incorporates two key factors in the optimality condition,
i.e., the maximization of the restoring speed (decoherence rate, $\Gamma\left(\omega\right)$)
and the minimization of the restoring target (equilibrium excited-state
population, $p_{e}^{\text{eq}}\left(\omega\right)$). Finally, the
optimal control
\begin{align}
\omega^{*}\left(t\right)= & \begin{cases}
\omega^{\text{min}}, & \omega^{*}\left(p_{e}\left(t\right)\right)<\omega^{\text{min}}\\
\omega^{\text{max}}, & \omega^{*}\left(p_{e}\left(t\right)\right)>\omega^{\text{max}}\\
\omega^{*}\left(p_{e}\left(t\right)\right), & \text{else}
\end{cases}\text{.}\label{EqOptCtrl}
\end{align}
is obtained by integrating the equation of motion, Eq. (\ref{EqEntropySt})
with the algebraic relationship Eq. (\ref{EqAlgRela}) and path constraint.

\begin{table*}[htbp]
\centering %
\begin{tabular}{lll}
\toprule 
Spectrum & Parameters & Platform\tabularnewline
\midrule 
\addlinespace
\parbox[c]{150pt}{%
\raggedright (a) Lorentzian:

${\displaystyle \Gamma_{\text{Lz}}\left(\omega\right)=\frac{g^{2}}{\kappa}\frac{\left(\frac{\kappa}{2}\right)^{2}}{\left(\omega-\omega_{r}\right)^{2}+\left(\frac{\kappa}{2}\right)^{2}}}$%
} & %
\parbox[c]{160pt}{%
\raggedright$\omega_{r}=2\pi\times\SI{5.4}{\giga\hertz},$

$\kappa=2\pi\times\SI{44}{\mega\hertz},g=2\pi\times\SI{107}{\mega\hertz}$%
} & %
\parbox[c]{190pt}{%
\raggedright Ref. \citep{houck2008controlling}; Purcell filter \citep{magnard2018fast,sete2014purcell,bronn2017fast};
two-level-system (TLS) defects \citep{klimov2018fluctuations,carroll2022dynamics};
IBM quantum computers \citep{abughanem2025ibm,kim2023evidence,carroll2022dynamics};
Google's Sycamore and Willow processors \citep{google2023suppressing,google2025quantum,klimov2018fluctuations};
the Ge/Si-based SC qubit from the Chinese Academy of Sciences \citep{zhuo2023hole}%
}\tabularnewline
\addlinespace
\parbox[c]{150pt}{%
\raggedright (b) Lorentzian with protection:

${\displaystyle \Gamma_{\text{prot}}\left(\omega\right)=\frac{4\kappa g^{2}\omega_{R}^{3}\left(\omega_{F}^{2}-\omega^{2}\right)^{2}}{\omega\left(\omega_{R}^{2}-\omega_{F}^{2}\right)^{2}\left(\omega_{R}^{2}-\omega^{2}\right)^{2}}}$%
} & %
\parbox[c]{160pt}{%
\raggedright$\omega_{F}=2\pi\times\SI{5.0}{\giga\hertz},\omega_{R}=2\pi\times\SI{6.5}{\giga\hertz},$

$\kappa=2\pi\times\SI{5}{\mega\hertz},g=2\pi\times\SI{150}{\mega\hertz}$%
} & %
\parbox[c]{190pt}{%
\raggedright Ref. \citep{bronn2015reducing}%
}\tabularnewline
\addlinespace
\parbox[c]{150pt}{%
\raggedright (c) Mixed:

${\displaystyle \Gamma_{\text{mix}}\left(\omega\right)=\frac{C_{\Phi}}{\omega^{0.9}}+C_{Q}\omega}$

${\displaystyle \ \ \ \ \ \ \ \ \ \ +\frac{C_{\text{Purcell}}\kappa^{2}}{\left(\omega-\omega_{r}\right)^{2}+\kappa^{2}}+C_{\text{other}}}$%
} & %
\parbox[c]{160pt}{%
\raggedright$C_{\Phi}=0.5,C_{Q}=0.001,C_{\text{Purcell}}=0.08,$
$\omega_{r}=8.27,\kappa=0.0015,C_{\text{other}}=0.02$

(in units of $2\pi\times\SI{}{\giga\hertz}$ for circular frequencies
and $\SI{}{\micro\second}$ for relaxation time)%
} & %
\parbox[c]{190pt}{%
\raggedright High-coherence flux qubit \citep{yan2016flux}%
}\tabularnewline
\addlinespace
\parbox[c]{150pt}{%
\raggedright (d) JQF:

${\displaystyle \Gamma_{\text{JQF}}\left(\omega\right)=\frac{1}{\tau_{0}+\frac{\tau\left(4\kappa_{j}\right)^{2}}{\left(\omega-\omega_{0}\right)^{2}+\left(4\kappa_{j}\right)^{2}}}}$%
} & %
\parbox[c]{160pt}{%
\raggedright$\omega_{0}=2\pi\times\SI{5.011}{\giga\hertz},$

$4\kappa_{j}=2\pi\times\SI{50.8}{\mega\hertz},$

$\tau_{0}=\SI{9.1}{\micro\second},\tau=\SI{98}{\micro\second}$%
} & %
\parbox[c]{190pt}{%
\raggedright Giant-atom-based Josephson quantum filter (JQF) \citep{hu2024engineering}%
}\tabularnewline
\bottomrule
\end{tabular}\caption{Four representative frequency-dependent decoherence-rate spectra,
$\Gamma\left(\omega\right)$, for SC-qubits. These spectra probe distinct
regimes of qubit--environment interaction and serve as benchmarks
for our switch--restore--switch scheme.}\label{TabFourSpectrum}
\end{table*}

\begin{figure}
\includegraphics{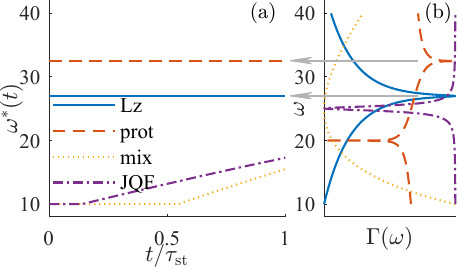} \caption{(a) Optimal control from Eq. (\ref{EqOptCtrl}). (b) Line shapes of four
representative decoherence-rate spectra, $\Gamma\left(\omega\right)$,
for SC-qubits. The blue solid, orange dashed, yellow dotted, and
purple dash-dotted lines represent the Lz, prot, mix, and JQF spectra,
respectively. The gray arrows indicate that the optimal control, $\omega^{*}$,
in (a) aligns with the peak of the spectrum in (b) for the Lz and prot
spectra. For the mix and JQF spectra, the restoring frequency initially
lies at the lower bound before gradually increasing. }\label{FigOmegaPlot}
\end{figure}

The optimal controls for four representative frequency-dependent decoherence-rate
spectra of superconducting qubits are shown in FIG.~\ref{FigOmegaPlot}(a),
and the corresponding line shapes are illustrated in FIG.~\ref{FigOmegaPlot}(b).
The spectra are listed in TABLE.~\ref{TabFourSpectrum}. For the Lz and prot
spectra, the optimal control remains essentially constant, $\omega^{*}\left(t\right)=\omega_{\text{st}}$,
which is exactly the frequency $\omega_{r/F}$ at which the decoherence
rate reaches its peak. Such constant optimal control naturally arises
in most solid-state qubit platforms operating at sub-$\SI{}{\kelvin}$
temperatures, such as superconducting qubits ($\sim\SI{10}{\milli\kelvin}$) \citep{wellstood1987low,devoret2013superconducting,gu2017microwave,kjaergaard2020superconducting,jiang2025advancements,jiang2024situ},
quantum-dot qubits ($\sim\SI{100}{\milli\kelvin}$) \citep{simmons2011tunable,harvey2022quantum,jansen2024high},
and trapped-ion qubits ($\sim\SI{10}{\milli\kelvin}$) \citep{de2022temperature}.
At these temperatures, the thermal-equilibrium population satisfies
$p_{e}^{\text{eq}}\left(\omega\right)\ll p_{e}$. Consequently, the
term $p_{e}^{\text{eq}}\left(\omega\right)$ in the optimality condition
becomes negligible, so the condition is overwhelmingly dominated
by $\Gamma\left(\omega\right)$. Under these circumstances, Eq. (\ref{EqAlgRela})
then reduces to the time-independent control, $\omega^{*}\left(t\right)=\omega_{\text{st}}$,
where the restoring frequency, $\omega_{\text{st}}$, is determined
solely by maximizing the decoherence rate,
\begin{align}
\omega_{\text{st}}= & \arg\max_{\omega\in\left[\omega^{\text{min}},\omega^{\text{max}}\right]}\Gamma\left(\omega\right)\text{.}
\end{align}
Such constant-frequency control is experimentally advantageous, as
it eliminates the need for real-time population measurements and feedback.

In the mix and JQF cases, the restoring frequency initially lies at
the lower bound, exploiting the maximal decoherence rate and thereby
decreasing the excited-state population, before gradually increasing
$\omega$ to satisfy the reset precision $\epsilon$. This highlights that in
those cases, the optimality condition in Eq. (\ref{EqAlgRela}) becomes
a trade-off between the maximization of the restoring speed and the
minimization of the restoring target. Notably, for the JQF spectrum,
the decoherence rate is largely insensitive to frequency far from
$\omega_{0}$, causing the restoring target to dominate the trade-off. 

\begin{figure}
\includegraphics{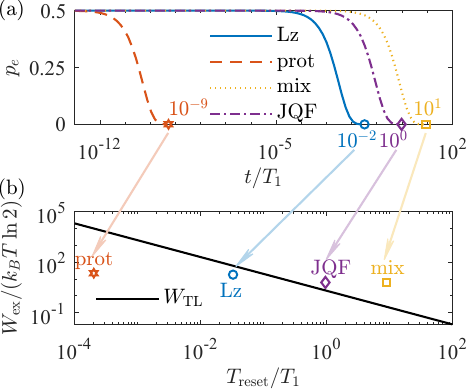} \caption{(a) Restoring process as a function of normalized time, $t/T_{1}$.
The blue solid, orange dashed, yellow dotted, and purple dash-dotted
lines represent the Lz, prot, mix, and JQF spectra, respectively. The end
of each restoring process is marked by a circle, hexagon, square,
or diamond, indicating the order of magnitude of the normalized
restoring time, $\tau_{\text{st}}/T_{1}$. The restoring time is shorter
than the coherence time except for the non-engineered mix spectrum. The
precision is set to $\epsilon=10^{-5}$. (b) Normalized extra work,
$W_{\text{ex}}/\left(k_{B}T\ln2\right)$, with respect to normalized
reset time, $T_{\text{reset}}/T_{1}$, for our switch--restore--switch
scheme, compared with the lower bound for a constant decoherence-rate
spectrum. The black line denotes the thermodynamic-length bound,
$W_{\text{TL}}/\left(k_{B}T\right)=1.4204/\left(T_{\text{reset}}/T_{1}\right)$,
for a constant-rate environment \citep{ma2022minimal}. The four lighter-colored
arrows, one for each spectrum, point from (a) to (b) and indicate $T_{\text{reset}}=\tau_{\text{st}}+2\tau_{\text{sw}}$.
For the prot and Lz spectra, our switch--restore--switch
scheme not only reduces the reset time but also lowers the extra work.}\label{FigEntropyT}
\end{figure}

For the four representative spectra, the dynamics of excited-state
population $p_{e}$ during the restoring process are shown in FIG.~\ref{FigEntropyT}(a).
The normalized restoring durations, $\tau_{\text{st}}/T_{1}$, are
$3.3\times10^{-2},5.3\times10^{-10},9.0\times10^{0}$, and $9.6\times10^{-1}$
for Lz, prot, mix, and JQF spectra, respectively. The Lz spectrum,
representing the most common instance and appearing in both non-engineered
and engineered settings \citep{houck2008controlling,sete2014purcell,bronn2015reducing,bronn2017fast,klimov2018fluctuations,google2023suppressing,zhuo2023hole,google2025quantum},
exhibits a favorable ratio due to its peak occurring at a relatively
high $\omega$. The engineered prot spectrum achieves a remarkably
small ratio, benefiting from the strong contrast of the decoherence
rate between the computation and restoring configurations. In contrast,
the non-engineered mix spectrum performs poorly, with a restoring
duration comparable to that of passive reset. The JQF spectrum, although
engineered, lacks sufficient contrast ($\Gamma\left(\omega_{\text{cp}}\right)/\Gamma\left(\omega_{\text{st}}\right)\sim0.1$),
leading to only modest improvement over passive reset.

Based on the observations above, we can highlight two guidelines for
spectral design. First, the decoherence rate should exhibit a
sufficiently high contrast between the computation and restoring configurations
($\Gamma\left(\omega_{\text{cp}}\right)$ and $\Gamma\left(\omega_{\text{st}}\right)$).
Second, $\Gamma\left(\omega\right)$ should display at
least an increasing trend with $\omega$,
ensuring that higher decoherence rates correspond to lower equilibrium
excited-state populations and thereby avoiding the
trade-off between the maximization of the restoring speed and the
minimization of the restoring target. Together, these features provide
concrete guidelines for spectral engineering aimed at achieving faster
and higher-fidelity qubit reset.

Adding the unavoidable switching duration, the shortest reset time
achieved by our switch--restore--switch scheme is $T_{\text{reset}}=\tau_{\text{st}}+2\tau_{\text{sw}}\sim\SI{20}{\nano\second}$
for the prot spectrum, limited primarily by the switching duration. This
reset time is substantially shorter than that of current active fast-reset
schemes that use frequency tunability \citep{reed2010fast,magnard2018fast,mcewen2021removing,zhou2021rapid,ding2025multipurpose} and is about $40\%$ of the
typical two-qubit gate time for tunable qubits (e.g., $\SI{50}{\nano\second}$ for the CZ (ad.) gate) \citep{chow2009randomized,kjaergaard2020superconducting}. Moreover,
the reset precision, set here to $\epsilon=10^{-5}$,
is well below the $\epsilon=10^{-3}$ level typical of high-fidelity fast
reset \citep{reed2010fast,zhou2021rapid,mcewen2021removing,johnson2022beating,yoshioka2023active,hu2024engineering},
thereby meeting the stringent requirements of future NISQ-era quantum
algorithms \citep{divincenzo2000physical,nielsen2002quantum,preskill2018quantum}.

As an information-erasing operation, the reset process is irreversible
in both the logical and thermodynamic senses \citep{landauer1961irreversibility,bennett1973logical,zurek1989thermodynamic,parrondo2015thermodynamics}.
Landauer’s principle establishes a fundamental lower limit on energetic
cost for reset \citep{landauer1961irreversibility}. The limit is
achieved only in the quasistatic limit, where the work equals the
free-energy change, while any finite-time implementation necessarily
requires additional dissipation \citep{diana2013finite,zulkowski2014optimal,proesmans2020finite,proesmans2020optimal,dago2021information,zhen2021universal,ma2022minimal,van2022finite,rolandi2023finite,huang2024qubit,pedersen2025optimal}.
FIG.~\ref{FigEntropyT}(b) shows the extra work of our switch--restore--switch
scheme operating in finite time, which is derived as \citep{EndMatter}
\begin{align}
W_{\text{ex}}= & \int_{0}^{\tau_{\text{st}}}\omega\Gamma\left(\omega\right)\left(p_{e}-p_{e}^{\text{eq}}\left(\omega\right)\right)\mathrm{d}t+T\Delta S\text{,}\label{EqExtraWork}
\end{align}
where $\Delta S=S\left(0.5\right)-S\left(\epsilon\right)$ is the
entropy reduction, with $S\left(p_{e}\right)=k_{B}\left(p_{e}\ln p_{e}+\left(1-p_{e}\right)\ln\left(1-p_{e}\right)\right)$.
Under the constant-control approximation and for very small $\epsilon$,
the extra work simplifies to $W_{\text{ex}}\approx0.5\omega_{\text{st}}-k_{B}T\ln2$,
which is independent of the reset duration and depends only on the
decoherence-rate spectrum. Numerically, the normalized extra work,
$W_{\text{ex}}/\left(k_{B}T\ln2\right)$, is $18.53,22.51,6.24$,
and $6.37$ for Lz, prot, mix, and JQF spectra, respectively. For
Lz and prot spectra, our switch--restore--switch scheme not only reduces
the reset time but also lowers the extra work, as evidenced by the
numerical values shown in FIG.~\ref{FigEntropyT}(b), compared with
the constant-rate bound from thermodynamic length theory \citep{ma2022minimal}.

In summary, we have developed a switch--restore--switch scheme that
minimizes qubit reset time through environmental spectral structure
and frequency switching, addressing a longstanding open question in quantum
information processing. In this scheme, the qubit is switched from
a low-decoherence frequency for computation to a high-decoherence
frequency for reset, followed by a return to the computational
setting. While the switching duration $\tau_{\text{sw}}$ is fixed
by qubit design, the restoring duration $\tau_{\text{st}}$ is minimized
within an optimal-control framework. We demonstrate the scheme
for superconducting qubits across four representative environmental spectra.
The reset time is shortened from $\gtrsim\SI{100}{\nano\second}$
to $\SI{20}{\nano\second}$, about $40\%$ of a typical two-qubit
gate time, while maintaining a reset precision as low as $10^{-5}$.
Moreover, we analyze the thermodynamic cost of finite-time reset
and find that, for the Lz and prot spectra, our switch--restore--switch scheme
not only reduces the reset time but also lowers the extra work.

We also discuss the robustness of our switch--restore--switch
scheme against common experimental imperfections. In particular, we
consider deviations in the initial population, initial coherence,
and control time. Their impact on the final state leads to errors
on the order of $\epsilon$. This robustness is further confirmed
by fidelity analysis with respect to the target final state, which
remains above $99.999\%$ over a wide range of deviations. Details
of the analysis are provided in the End Matter \citep{EndMatter}.
\begin{acknowledgments}
This work is supported by the National Natural Science Foundation
of China (Grant Nos. 12088101, U2230203, and U2330401), and the Quantum
Science and Technology-National Science and Technology Major Project
(Grant No. 2023ZD0300700).
\end{acknowledgments}

\bibliographystyle{apsrev4-2}
\bibliography{QubitResetTime}

\balancecolsandclearpage

\appendix

\section*{End Matter}

\textit{The Lindbladian superoperator and the equation of motion.} The
Lindblad master equation \citep{breuer2002theory,wang2021nonadiabatic,naeij2025open}
is $\dot{\rho}_{S}=-\mathrm{i}\left[H_{S},\rho_{S}\right]+\mathcal{L}\left[\rho_{S}\right]$
, where 
\begin{align}
\mathcal{L}\left[\rho_{S}\right]= & -\gamma\left(\omega\right)\left(n\left(\omega\right)+1\right)\left(\frac{1}{2}\left\{ \sigma_{+}\sigma_{-},\rho_{S}\right\} -\sigma_{-}\rho_{S}\sigma_{+}\right)\nonumber \\
 & -\gamma\left(\omega\right)n\left(\omega\right)\left(\frac{1}{2}\left\{ \sigma_{-}\sigma_{+},\rho_{S}\right\} -\sigma_{+}\rho_{S}\sigma\right)
\end{align}
is the Lindbladian superoperator, where $\gamma\left(\omega\right)$
is the decay rate, and $n\left(\omega\right)=1/2\left(\coth\left(\beta\omega/2\right)-1\right)$
is the average occupation number of the environment. $\left\{ A,B\right\} =AB+BA$
denotes the anticommutator of operators $A$ and $B$. $\sigma_{+}=\left|e\right\rangle \left\langle g\right|,\sigma_{-}=\left|g\right\rangle \left\langle e\right|,\sigma_{z}=\left|e\right\rangle \left\langle e\right|-\left|g\right\rangle \left\langle g\right|$
are the Pauli operators, with $\left|g\right\rangle$ and $\left|e\right\rangle$
denoting the ground and excited states of the qubit Hamiltonian, respectively. Introducing
the total decoherence rate, $\Gamma\left(\omega\right)=\gamma\left(\omega\right)\left(2n\left(\omega\right)+1\right)$,
and the equilibrium excited-state population, $p_{e}^{\text{eq}}\left(\omega\right)=n\left(\omega\right)/\left(2n\left(\omega\right)+1\right)$,
the matrix equation can be written in terms of its elements as 
\begin{subequations}
\begin{align}
\dot{p}_{e}= & -\Gamma\left(\omega\right)p_{e}+\Gamma\left(\omega\right)p_{e}^{\text{eq}}\left(\omega\right)\text{,}\label{EqMain}\\
\dot{p}_{r}= & -\frac{1}{2}\Gamma\left(\omega\right)p_{r}+\omega p_{i}\text{,}\label{EqMinor1}\\
\dot{p}_{i}= & -\frac{1}{2}\Gamma\left(\omega\right)p_{i}-\omega p_{r}\text{.}\label{EqMinor2}
\end{align}
\label{EqMainAll}\\
\end{subequations}
where $p_{e}=\left\langle e\right|\rho_{S}\left|e\right\rangle ,p_{r}=\Re\left(\left\langle e\right|\rho_{S}\left|g\right\rangle \right),p_{i}=\Im\left(\left\langle e\right|\rho_{S}\left|g\right\rangle \right)$.

\textit{The details of Pontryagin's minimum principle.} Pontryagin’s
minimum principle provides necessary conditions for time-optimal control
problems, whose cost function is the final time $\tau=\int_{0}^{\tau}\ \mathrm{d}t$,
with motion equation $\dot{x}=f\left(x,u\right)$, where $x$ denotes
the state and $u$ denotes the control. These conditions are expressed
in terms of the control Hamiltonian, $\mathcal{H}\left(x,u\right)=1+\lambda f\left(x,u\right)$,
where the Lagrange multiplier $\lambda$ is the costate. These
conditions can be stated as follows \citep{pontryagin1987mathematical}:
\begin{enumerate}
\item The optimal control $u^{*}$ minimizes $\mathcal{H}$ pointwise in
time, ensuring that at each instant the system follows the minimal-cost
path.
\item The state $x$ and costate $\lambda$ evolve according to Hamilton’s
canonical equations, which describe how the system changes over time.
\item For a time-optimal control problem with free final time, the
transversality condition, $\mathcal{H}\left(\tau\right)=0$, imposes
a constraint at the final time, linking the terminal condition of
the state and costate.
\end{enumerate}
Together, these conditions not only determine the optimal control
but also provide qualitative information about the costate variables,
such as their monotonicity or sign, which can be used to infer the
structure of the solution.

In the setting of this Letter ($x\to p_{e},u\to\omega,\tau\to\tau_{\text{st}}$),
the costate obeys a linear equation of the form $\dot{\lambda}=\Gamma\left(\omega\right)\lambda$,
whose solution reads $\lambda\left(t\right)=\lambda\left(\tau_{\text{st}}\right)\exp\left(-\int_{t}^{\tau_{\text{st}}}\Gamma\left(\omega\left(s\right)\right)\ \mathrm{d}s\right)$.
The transversality condition fixes the terminal value as
\begin{align}
\lambda\left(\tau_{\text{st}}\right)= & \frac{1}{\left(\Gamma\left(\omega\left(\tau_{\text{st}}\right)\right)\left(p_{e}\left(\tau_{\text{st}}\right)-p_{e}^{\text{eq}}\left(\omega\left(\tau_{\text{st}}\right)\right)\right)\right)}\text{.}
\end{align}
Provided the denominator is finite and nonvanishing, this expression
determines the sign of $\lambda\left(\tau_{\text{st}}\right)$. Since
the exponential factor is strictly positive, the sign of $\lambda\left(t\right)$
is preserved along the trajectory, implying $\lambda\left(t\right)>0$
for all $t$. This reflects the general structure of Pontryagin’s
framework, where the costate evolves multiplicatively and inherits
its sign from the terminal condition imposed by the transversality
condition.

\textit{The robustness against common errors in practical experiments.}
Our switch--restore--switch scheme is not only fast but also robust
against common experimental errors. To assess this robustness,
we examine three types of deviations as follows. (1) Population
deviation: the initial state is $\rho_{S}\left(0\right)=\mathrm{diag}\left(p,1-p\right)$,
where $p\in\left[0,1\right]$. (2) Coherence deviation: the initial
state is $\rho_{S}\left(0\right)=\left(\begin{array}{cc}
\frac{1}{2} & c\\
c^{*} & \frac{1}{2}
\end{array}\right)$, where $c\in\mathbb{C}$ with $\left|c\right|\in\left[0,0.5\right]$
because of the positive semidefiniteness of the density matrix. (3) Control-time
deviation: the control time is $t_{f}=\tau+\delta\tau$.

\begin{figure}
\includegraphics{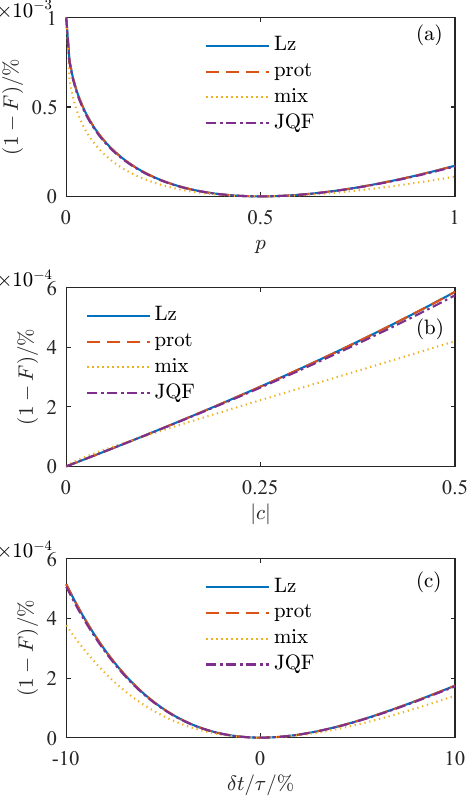} \caption{The fidelity, $F$, of our restoring process under three types of
deviations: (a) population deviations, (b) coherence deviations, and (c) control-time
deviations. The blue solid, orange dashed, yellow dotted, and purple dash-dotted lines represent
the Lz, prot, mix, and JQF spectra, respectively. In all cases,
$F>99.999\%$ over substantial deviation ranges. The reset
precision is set to $\epsilon=10^{-5}$.}\label{FigRobustness}
\end{figure}

By defining the decoherence factor as $\eta\left(t\right)=\exp\left(-\int_{0}^{t}\Gamma\left(s\right)\ \mathrm{d}s\right)$
with $\eta\left(\tau\right)<2\epsilon$, the robustness to
population, coherence, and control-time deviations is characterized,
respectively, by
\begin{subequations}
\begin{align}
\frac{\partial p_{e}\left(\tau\right)}{\partial p}= & \eta\left(\tau\right)\text{,}\\
\frac{\partial\left|\rho_{eg}\left(\tau\right)\right|}{\partial\left|c\right|}= & \eta\left(\tau\right)\text{,}\\
\left|\frac{\partial p\left(\tau+\delta\tau\right)}{\partial\left(\Gamma\delta\tau\right)}\right|= & \epsilon\eta\left(\delta\tau\right)\text{,}
\end{align}
\\
\end{subequations}
which are derived from Eq. (\ref{EqMainAll}) and are of order $\epsilon$.
Quantitatively, the deviation of the final state is quantified by
the fidelity, $F\left(\rho_{\text{tg}},\rho\right)=\left(\mathrm{Tr}\sqrt{\sqrt{\rho_{\text{tg}}}\rho\sqrt{\rho_{\text{tg}}}}\right)^{2}$,
where $\rho_{\text{tg}}=\mathrm{diag}\left(1-\epsilon,\epsilon\right)$
is the target final state. Numerical results for the population,
coherence, and control-time deviations are shown in Fig. \ref{FigRobustness}.
In all cases, $F>99.999\%$ over substantial deviation ranges,
confirming the robustness of our scheme against
these deviations.

\textit{The extra work done in our switch--restore--switch scheme.}
The work in our switch--restore--switch scheme can be written as $W=W_{\text{sw1}}+W_{\text{st}}+W_{\text{sw2}}$,
where
\begin{align}
W_{\text{sw1}}= & \left(\omega\left(0\right)-\omega_{\text{cp}}\right)\left(p_{e}\left(0\right)-0.5\right)\text{,}\\
W_{\text{sw2}}= & \left(\omega_{\text{cp}}-\omega\left(\tau_{\text{st}}\right)\right)\left(p_{e}\left(\tau_{\text{st}}\right)-0.5\right)
\end{align}
are the work contributions from the two switching stages, and $W_{\text{st}}=\int_{0}^{\tau_{\text{st}}}\dot{\omega}p_{e}\mathrm{d}t$.
Integrating by parts and using the dynamics in Eq. (\ref{EqMain}), we obtain
\begin{align}
W_{\text{st}}= & \omega_{\text{cp}}\left(p_{e}\left(\tau_{\text{st}}\right)-p_{e}\left(0\right)\right)+\int_{0}^{\tau_{\text{st}}}\omega\Gamma\left(\omega\right)\left(p_{e}-p_{e}^{\text{eq}}\left(\omega\right)\right)\mathrm{d}t\text{.}
\end{align}
The change in free energy during reset is $\Delta F=\Delta U-T\Delta S$,
where $\Delta U=\omega_{\text{cp}}\left(p_{e}\left(\tau_{\text{st}}\right)-p_{e}\left(0\right)\right)$
and $\Delta S=S\left(0.5\right)-S\left(\epsilon\right)$ with $S\left(p_{e}\right)=k_{B}\left(p_{e}\ln p_{e}+\left(1-p_{e}\right)\ln\left(1-p_{e}\right)\right)$.
Consequently, the extra work $W_{\text{ex}}=W-\Delta F$ is given by Eq. (\ref{EqExtraWork}).
\end{document}